\documentclass[12pt,a4paper]{article}
\usepackage{epsfig}
\begin{document}
\title{Lepton flavor violating semileptonic $\tau$ decays
$\tau\rightarrow lP(V)$ in a topcolor scenario
\\
\hspace*{-0.8cm}  }

\author{Chong-Xing Yue, Li-Hong Wang, Wei Ma\\
{\small  Department of Physics, Liaoning Normal University, Dalian
116029, China}\thanks{E-mail:cxyue@lnnu.edu.cn}\\}
\date{\today}

\maketitle
\begin{abstract}

The contributions of the neutral top-pion $\pi_{t}^{0}$ and the
non-universal gauge boson $Z'$ predicted by topcolor scenario to the
lepton flavor violating $(LFV)$ semileptonic $\tau$ decays
$\tau\rightarrow lP(V)$ $(P=\pi^{0},\ \eta,\ \eta'\ $ and $\
V=\rho^{0},\ \phi)$ are discussed. We find that the contributions of
$Z'$ to these decay processes are generally larger than those from
$\pi_{t}^{0}$. $\pi_{t}^{0}$ can only make the value of the
branching ratio $Br(\tau\rightarrow lP)$ in the range of
1$\times10^{-11}$ $\sim$ 1$\times10^{-16}$, which is far below the
sensitivity of foreseeable experiments. With reasonable values of
the free parameters, the non-universal gauge boson $Z'$ can make the
value of the branching ratio $Br(\tau^{-}\rightarrow \mu^{-}\phi)$
reach 1$\times10^{-7}$, which might approach the observable
threshold of near-future experiments.

\vspace{2.0cm} \noindent
 {\bf PACS number(s)}:12.60.Cn, 13.35.Dx, 14.80.Cp.

\end{abstract}

\newpage
\noindent{\bf I. Introduction}

The flavor physics of quarks and leptons is one of the most
important issue of current particle physics. Over the past decade
years, the most surprising development in flavor physics is
observation of neutrino oscillation, which can be seen as the first
experimental clue for new physics beyond the standard model ($SM$)
[1]. Observation of neutrino masses also provides the
 evidence for flavor violating in the lepton sector and gives the
possibility of lepton flavor violating ($LFV$) among the charged
leptons. It is well known that the tree-level $LFV$ processes are
absent in the $SM$, due to unitary of the leptonic analog of
Cabibbo-Kobayashi-Maskawa ($CKM$) mixing matrix and the masslessness
of three neutrinos. Thus, observation of the $LFV$ processes would
be a clear signature of new physics beyond the $SM$. This fact has
led to a great amount of theoretical effect for revealing the
underlying new physics in the leptonic flavor sector.

The $\tau$ lepton is the most heavy particle in the leptonic sector
of the $SM$, which is much more sensitive than the lepton $e$ or
$\mu$ to new physics related to the flavor and mass-generation
problems [2]. The leptonic or semileptonic character of $\tau$
decays provides a clean laboratory to test the structure of the weak
currents and the universality of their couplings to the gauge
bosons. Furthermore, its semileptonic decay is ideal tool for
studying strong interaction effects in very clean conditions.
Moreover, the sensitivity of probing the $LFV$ semileptonic $\tau$
decays have been enhanced to $\cal{O}$$(10^{-7})$ [3]. Thus, in the
framework of some popular models beyond the $SM$, studying the
semileptonic $\tau$ decays related $LFV$ is very interesting and
needed.

The discovery of $LFV$ in the neutrino oscillation experiments has
opened a new era for flavor physics in the leptonic sector, where
one can study the possible signatures of new physics via some $LFV$
processes. The effects of new physics on the $LFV$ semileptonic
$\tau$ decays, such as $\tau\rightarrow lP$, $\tau\rightarrow lV$,
and $\tau\rightarrow lPP$, have been extensively studied in
Refs.[4,5,6,7], where $P(=\pi^{0},\eta,\eta')$ and
$V(=\rho^{0},\phi)$ represent the pseudoscalar meson and vector
meson, respectively. It has been shown that these decay processes
are very sensitive to the new physics effects and the values of the
branching ratios for some of these processes might be enhanced to
the experimental interesting ranges. The constraints on the free
parameters of some specific models beyond the $SM$ have been
obtained.

To completely avoid the problems arising from the elementary Higgs
field in the $SM$, various kinds of dynamical electroweak symmetry
breaking $(EWSB)$ models have been proposed, among which topcolor
scenario is attractive because it can explain the large top quark
mass and provide a possible $EWSB$ mechanism [8]. Almost all of this
kind of models propose that the underlying interactions, topcolor
interactions, should be flavor non-universal. When one writes the
non-universal interactions in the mass eigenbasis, it can induce the
tree-level flavor changing ($FC$) couplings, which can generate rich
phenomenology.

A common feature of the topcolor models, such as topcolor-assisted
technicolor ($TC2$) models [9], flavor-universal $TC2$ models [10],
top see-saw models [11] and top flavor see-saw models [12], is that
the physical top-pions ($\pi^{0,\pm}_{t}$) and non-universal gauge
boson $Z'$ are predicted. These new particles treat the third
generation fermions differently from those in the first and second
generations and thus can lead to the tree-level $FC$ couplings. The
aim of this paper is to study the contributions of these new
particles to the $LFV$ semileptonic $\tau$ decays $\tau\rightarrow
lP$ and $\tau\rightarrow lV$ and see whether the values of their
branching ratios can be significantly enhanced.

To predigestion our calculation, we will give our numerical results
in the context of the $TC2$ models. In the next section, we will
briefly summarize the relevant flavor-diagonal ($FD$) and $FC$
coupling expressions of the new particles (the neutral top-pion
$\pi^{0}_{t}$ and non-universal gauge boson $Z'$) predicted by the
$TC2$ models. The contributions of $\pi^{0}_{t}$ and $Z'$ to the
$LFV$ semileptonic $\tau$ decays $\tau\rightarrow lP$ and
$\tau\rightarrow lV$ are calculated in Sec.III and Sec.IV,
respectively. Section $V$ contains our conclusions.

\noindent{\bf II. The relevant couplings of the neutral top-pion
$\pi^{0}_{t}$ and the non-universal \hspace*{0.55cm} gauge boson
$Z'$}

In topcolor scenario [8], topcolor interactions, which are not
flavor-universal and mainly couple to the third generation fermions,
generally generate small contributions to $EWSB$ and give rise to
the main part of the top quark mass. Thus, the top-pions
($\pi^{0,\pm}_{t}$) have large Yukawa couplings to the third
generation fermions, and can induce the new $FC$ couplings. In the
$TC2$ models, the $FD$ and $FC$ couplings of the neutral top-pion
$\pi^{0}_{t}$ to light fermions, which are related our calculation,
can be written as [8,9,13,14]:
\begin{equation}
\frac{m_{f}}{\nu}\bar{f}\gamma^{5}f\pi^{0}_{t}+\frac{m_{\tau}}{\nu}
K\bar{\tau}\gamma^{5}l\pi^{0} _{t},
\end{equation}
where $\nu=\nu_{W}/\sqrt{2}\approx174GeV$, $f$ represents the light
quark $(u,\ d,\ c, $ or $s)$, and $l$ represents the first (second)
generation lepton $e(\mu)$. $K$ is the lepton flavor mixing factor
between the third- and the first- or second- generation leptons.
Certainly, there is also the $FC$ coupling $\pi^{0}_{t}\bar{\mu}e$.
However, the topcolor interactions only contact with the
third-generation fermions, and thus the flavor mixing between the
first- and second- generation fermions is very small, which can be
ignored [15].

An inevitable feature of topcolor scenario is that the $SM$ gauge
groups are extended at energy well above the weak scale. Breaking of
the extended groups to their diagonal subgroups produces the
non-universal massive gauge boson $Z'$ [16]. This kind of new
particles generally couple primarily to the third generation
fermions and have large tree-level $FC$ couplings.

The $FD$ couplings of $Z'$ to fermions, which are related our
calculation, can be written as [8,9,17]:
\begin{eqnarray}
L^{FD}_{Z'}&=&-\sqrt{4\pi K_{1}}\{
Z'_{\mu}[\frac{1}{2}\bar{\tau}_{L}\gamma^{\mu}\tau_{L}
-\bar{\tau}_{R}\gamma^{\mu}\tau_{R}]-\tan^{2}
\theta'Z'_{\mu}[\frac{1}{6}\bar{c}_{L}\gamma^{\mu}c_{L}+
\frac{2}{3}\bar{c}_{R}\gamma^{\mu}c_{R}\nonumber\\
&+&\frac{1}{6}\bar{s}_{L}\gamma^{\mu}s_{L}
-\frac{1}{3}\bar{s}_{R}\gamma^{\mu}s_{R}-\frac{1}{2}
\bar{\mu}_{L}\gamma^{\mu}\mu_{L}-\bar{\mu}_{R}\gamma^{\mu}\mu_{R}+\frac{1}{6}\bar{u}_{L}\gamma^{\mu}
u_{L}+\frac{1}{6}\bar{d}_{L}\gamma^{\mu}d_{L}\nonumber\\
&+&\frac{2}{3}\bar{u}_{R}\gamma^{\mu}u_{R}
-\frac{1}{3}\bar{d}_{R}\gamma^{\mu}d_{R}-\frac{1}{2}\bar{e}_{L}\gamma^{\mu}e_{L}-\bar{e}_{R}\gamma^{\mu}e_{R}]\},
\end{eqnarray}
where $K_{1}$ is the coupling constant and $\theta'$ is the mixing
angle with $\tan \theta'=\frac{g_{1}}{\sqrt{4\pi K_{1}}}$. $g_{1}$
is the ordinary hypercharge gauge coupling constant. To obtain the
top quark condensation and not form a $b\bar{b}$ condensation, there
must be $\tan \theta'\ll1$ [9,10]. In above equation, we have
assumed that there is no mixing between the $SM$ gauge boson $Z$ and
the non-universal gauge boson $Z'$. The $FC$ couplings of $Z'$ to
leptons can be written as [17,18]:
\begin{eqnarray}
L^{FC}_{Z'}=\frac{1}{2}g_{1}K'Z'_{\mu}[\bar{\tau}_{L}\gamma^{\mu}\mu_{L}+
2\bar{\tau}_{R}\gamma^{\mu}\mu_{R}+\bar{\tau}_{L}\gamma^{\mu}e_{L}+2\bar{\tau}_{R}\gamma^{\mu}e_{R}],
\end{eqnarray}
where $K'$ is the lepton flavor mixing factor. Since the
non-universal gauge boson $Z'$ treats the fermions in the third
generation differently from those in the first and second generation
and treats the fermions in the first generation same as those in the
second generation, so we have assumed $K'_{\tau\mu}=K'_{\tau e}=K'$
in above equation. In this case, the contributions of $Z'$ to the
$LFV$ semileptonic $\tau$ decays $\tau\rightarrow\mu P(V)$ are
approximately equal to those for the decays $\tau\rightarrow e
P(V)$.

Integrating out the non-universal gauge bosons $Z'$, Eq.(2) and
Eq.(3) can give rise to the effective four fermion couplings
$\tau\mu qq\ (q=u,\ d,\ c$, and $s)$:
\begin{eqnarray}
L_{4f} &=& -\frac{\pi K_{1}K'\tan^{3}
\theta'}{M^{2}_{Z'}}(\bar{\tau}_{L}\gamma^{\mu}\mu_{L}+
2\bar{\tau}_{R}\gamma^{\mu}\mu_{R})[\frac{1}{6}(\bar{c}_{L}\gamma_{\mu}c_{L}
+\bar{s}_{L}\gamma_{\mu} s_{L}+\bar{u}_{L}\gamma^{\mu}u_{L}\nonumber\\
 &+& \bar{d}_{L}\gamma_{\mu}d_{L})
+\frac{2}{3}(\bar{c}_{R}\gamma_{\mu}c_{R}+\bar{u}_{R}\gamma^{\mu}u_{R})
-\frac{1}{3}(\bar{s}_{R}\gamma^{\mu}s_{R}+\bar{d}_{R}\gamma^{\mu}d_{R})].
\end{eqnarray}
Where $M_{Z'}$ is the mass of the non-universal gauge boson $Z'$.

In the following sections, we will use above coupling expressions to
calculate the branching ratios $Br(\tau\rightarrow \mu V)$ and
$Br(\tau\rightarrow \mu P)$, and compare our numerical results with
the corresponding experimental upper limits given in Table 1[3].
\begin{center}
{
\begin{small}
\begin{tabular}{|c|c|}
\hline Decay\ Process&Current \ Upper \ Limit(90\% \ C.L.)  \\
\hline
$\tau^{-}\rightarrow \mu^{-}\pi^{0}$&$4.1\times10^{-7}$ \\
$\tau^{-}\rightarrow \mu^{-}\eta$&$2.3\times10^{-7}$\\
$\tau^{-}\rightarrow \mu^{-}\eta^{'}$&$4.7\times10^{-7}$\\
$\tau^{-}\rightarrow \mu^{-}\rho^{0}$&$2.0\times10^{-7}$\\
$\tau^{-}\rightarrow\mu^{-}\phi$&$7.7\times10^{-7}$\\
\hline\end{tabular}
\end{small} }\end{center}
\hspace{0.3cm} Table 1: Current experimental upper limits on the
branching ratios $Br(\tau\rightarrow\mu P(V))$. \hspace*{1.9cm}
\vspace*{0.5cm}

\noindent{\bf III. The neutral top-pion $\pi^{0}_{t}$ and the $LFV$
semileptonic $\tau$ decay $\tau\rightarrow\mu P$}

It is well known that the $LFV$ semileptonic decay $\tau\rightarrow
lS$ ($S$ denotes a scalar meson) can only be generated by the scalar
current, the decay $\tau\rightarrow lV$ ($V$ denotes a vector meson)
can only be generated by the vector current, while the decay
$\tau\rightarrow lP$ ($P$ denotes a pseudoscalar meson) can be
induced by the axial-vector or pseudoscalar currents. The neutral
top-pion $\pi^{0}_{t}$ is the CP-odd pseudoscalar particle, thus it
can only have contributions to the decay $\tau\rightarrow lP$ via
the $FC$ lepton couplings and the $FD$ light quark couplings.

To calculate the branching ratios of the $LFV$ semileptonic $\tau$
decays $\tau^{-}\rightarrow\mu^{-}\pi^{0}$, $\mu^{-}\eta$, and
$\mu^{-}\eta'$, we write the relevant pseudoscalar matrix elements
as [4]:
\begin{eqnarray}
\langle0|\bar{u}\gamma^{5}u|\pi^{0}(p)\rangle &=&
-\langle0|\bar{d}\gamma^{5}d|\pi^{0}(p)\rangle
=\frac{i}{\sqrt{2}}\frac{m_{\pi}^{2}}{m_{u}+m_{d}}F_{\pi},\\
\langle0|\bar{s}\gamma^{5}s|\eta_{8}(p)\rangle &=&
-i\sqrt{6}F_{\eta}^{8}\frac{m_{\eta_{8}}^{2}} {m_{u}+m_{d}+4m_{s}},\\
\langle0|\bar{s}\gamma^{5}s|\eta'_{8}(p)\rangle &=&
-i\sqrt{6}F_{\eta'}^{8}
\frac{m^{2}_{\eta'_{8}}}{m_{u}+m_{d}+4m_{s}}.
\end{eqnarray}
Where $F_{\pi}$, $F^{8}_{\eta}$, and $F^{8}_{\eta'}$ are the decay
constants of the pseudoscalar mesons $\pi^{0}$, $\eta_{8}$, and
$\eta'_{8}$, respectively.

Using Eq.(1), Eq.(5), Eq.(6), and Eq.(7), we can give the
expressions of the branching ratios $Br(\tau^{-}\rightarrow
\mu^{-}\pi^{0})$, $Br(\tau^{-}\rightarrow\mu^{-}\eta)$, and
$Br(\tau^{-}\rightarrow\mu^{-}\eta')$ generated by the neutral
top-pion $\pi_{t}^{0}$ as:
\begin{eqnarray}
Br(\tau^{-}\rightarrow \mu^{-}\pi^{0})&=&\frac{6K^{2}}{\cos^{2}
\theta_{c}}(\frac{m_{\pi}}{M_{\pi_{t}}})^{4}(\frac{m_{u}-m_{d}}{m_{u}+m_{d}})^{2}Br
(\tau^{-}\rightarrow \nu_{\tau}\pi^{-}), \\Br(\tau^{-}\rightarrow
\mu^{-}\eta)&=&\frac{18K^{2}}{\cos^{2}
\theta_{c}}(\frac{F_{\eta}}{F_{\pi}})^{2}(\frac{m_{\eta}}{M_{\pi_{t}}})^{4}(\frac{m_{u}+m_{d}-2m_{s}}
{m_{u}+m_{d}+4m_{s}})
^{2}Br(\tau^{-}\rightarrow \nu_{\tau}\pi^{-}), \\
Br(\tau^{-}\rightarrow \mu^{-}\eta')&=&\frac{18K^{2}}{\cos^{2}
\theta_{c}}(\frac{F_{\eta^{'}}}{F_{\pi}})^{2}(\frac{m_{\eta^{'}}}{M_{\pi_{t}}})^{4}(\frac{m_{u}+m_{d}-2m_{s}}
{m_{u}+m_{d}+4m_{s}})
^{2}Br (\tau^{-}\rightarrow \nu_{\tau}\pi^{-}).
\end{eqnarray}
Where the meson decay constants are defined as:
$F_{\eta}=F^{8}_{\eta}-\frac{1}{\sqrt{2}}F^{0}_{\eta}$ and
$F_{\eta'}=F^{8}_{\eta'}+\frac{1}{\sqrt{2}}F^{0}_{\eta'}$.
$\theta_{c}$ is the Cabibbo angle. $M_{\pi_{t}}$ represents the mass
of the physical top-pions ($\pi_{t}^{0,\pm}$), its value remains
subject to large uncertainty [8]. However, it has been shown that
its value is generally allowed to be in the range of a few hundred
$GeV$ depending on the models [19]. In our numerical estimation, we
will take $M_{\pi_{t}}$ as a free parameter and assume that it is in
the range of $150GeV\sim400GeV$.

Certainly, the neutral top-pion $\pi_{t}^{0}$ can also generate
contributions to the $LFV$ semileptonic decay $\tau^{-}\rightarrow
\mu^{-} P$ via the $Z$ penguin diagrams, $i. \ e.$ the effective
process $\tau^{-}\rightarrow \mu^{-}Z^{*}\rightarrow
\mu^{-}f\bar{f}$. However, compared with those from $\pi_{t}^{0}$
exchange at the tree-level, the contributions are much small. So, we
do not consider the one-loop contributions of $\pi_{t}^{0}$ to the
decay $\tau^{-}\rightarrow \mu^{-} P$ in this paper.

\begin{figure}[htb]
\begin{center}
\epsfig{file=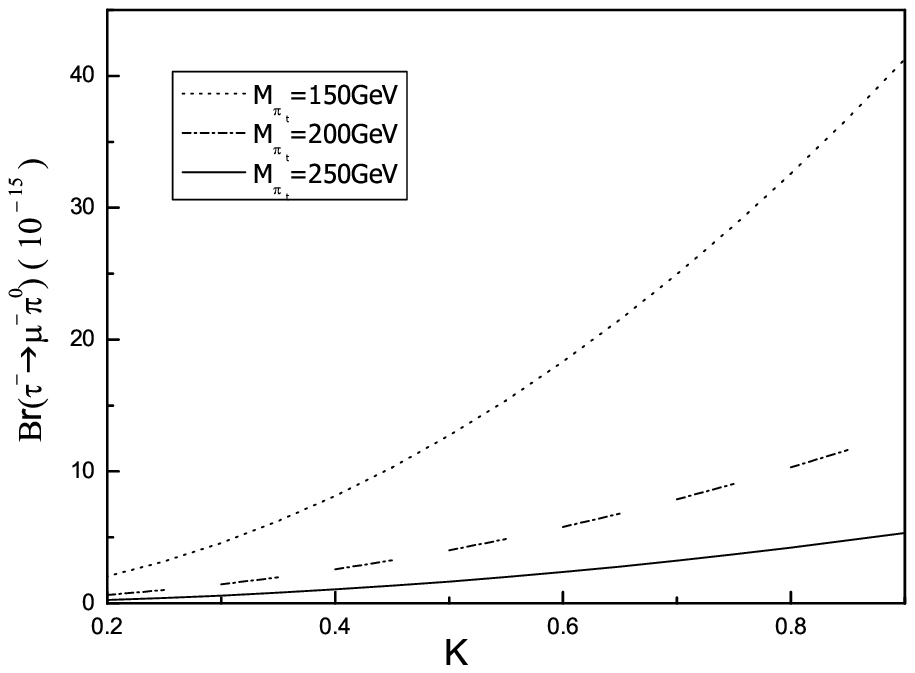,width=220pt,height=200pt} \put(-280,3) {Fig.1:
The branching ratio $Br(\tau^{-}\rightarrow \mu^{-}\pi^{0})$ as a
function of the mixing } \put(-250,-10) {parameter $K$ for three
values of the mass parameter $M_{\pi_{t}}$.} \hspace{-0.5cm}
\hspace{10cm}\vspace{0.5cm}
\epsfig{file=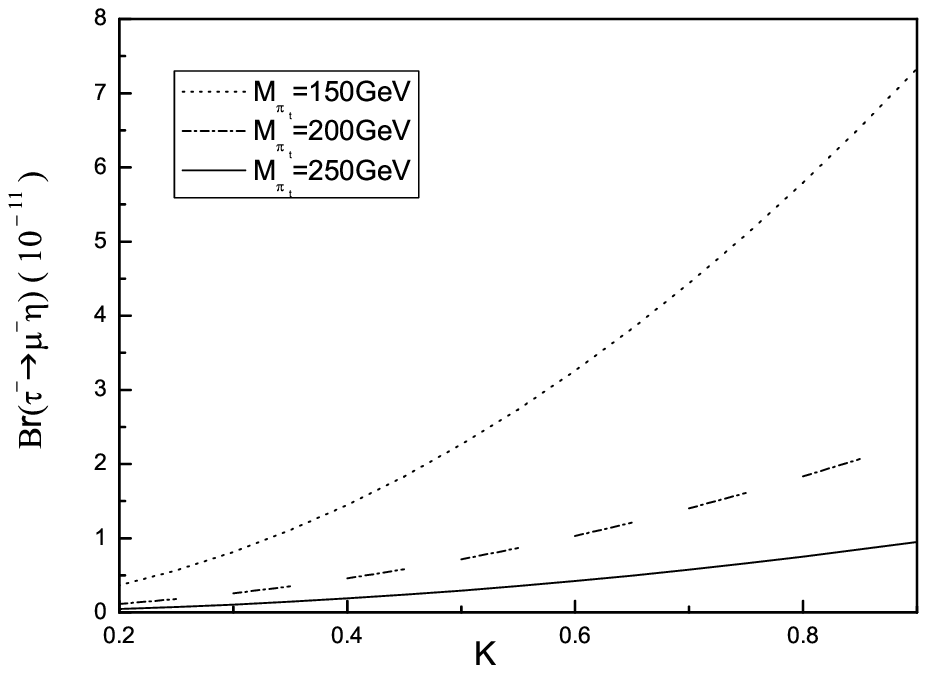,width=220pt,height=200pt}\hspace{-0.5cm}\hspace{0cm}
\vspace{-0.25cm} \put(-230,-8){Fig.2: Same as Fig.1 but for
$Br(\tau^{-}\rightarrow \mu^{-}\eta)$. }
\epsfig{file=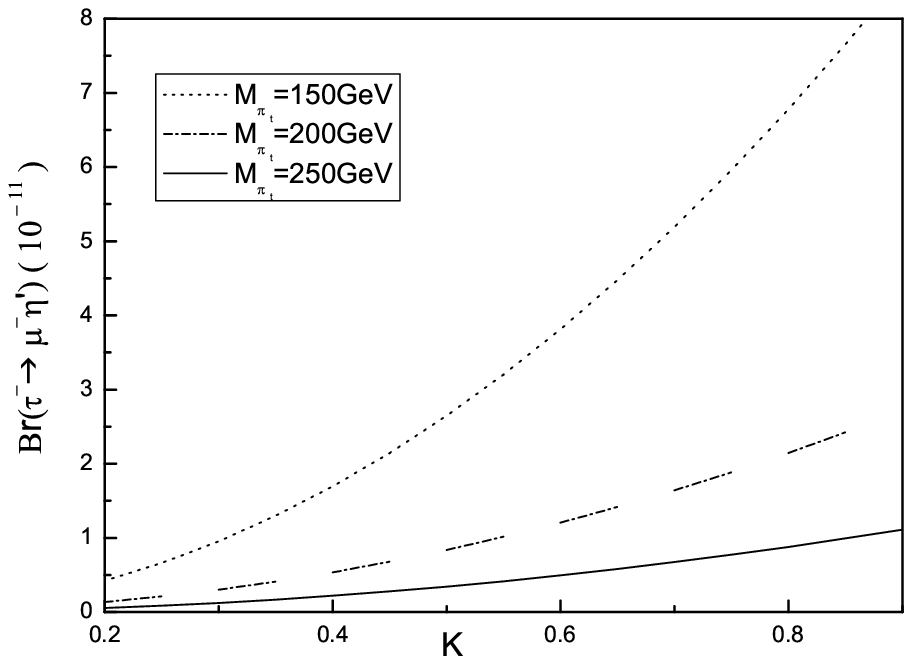,width=220pt,height=200pt} \put(-210,-8){Fig.3:
Same as Fig.1 but for $Br(\tau^{-}\rightarrow \mu^{-}\eta')$.
}\hspace{-0.5cm}
 \hspace{10cm}\vspace{-1cm}

\label{ee}
\end{center}
\end{figure}
\vspace{5cm}

Except the free parameter $M_{\pi_{t}}$, the branching ratio
$Br(\tau^{-}\rightarrow \mu^{-}P)$ depends on the mixing factor $K$.
Topcolor scenario has not given any prediction about its value. In
general, the experimental data about observables, such as $\mu$
anomalous magnetic moment $a_{\mu}$, the branching ratios
$Br(\tau\rightarrow l_{i}\gamma)$ and $Br(\tau\rightarrow
l_{i}l_{j}l_{k})$, can give constraints on the values of the free
parameter $K$. However, although the neutral top-pion $\pi_{t}^{0}$
can generate significant contributions to the $LFV$ processes
$\tau\rightarrow l_{i}\gamma$ and $\tau\rightarrow l_{i}l_{j}l_{k}$
via the $FC$ couplings, the current experimental upper limits on
$Br(\tau\rightarrow l_{i}\gamma)$ and $Br(\tau\rightarrow
l_{i}l_{j}l_{k})$ can not give severe constraints on the mixing
factor $K$ [20]. Thus, in this paper, we will assume that the value
of the mixing factor $K$ is in the range of $0.1\sim0.9$.

In our numerical estimation, we will take $F_{\pi}=131MeV$,
$F^{8}_{\eta}\approx1.2F_{\pi}$, $F^{0}_{\eta}\approx0.2F_{\pi}$,
$F^{8}_{\eta'}\approx-0.45F_{\pi}$,
$F^{0}_{\eta'}\approx1.15F_{\pi}$ [21]. The other $SM$ input
parameters are taken as: $Br (\tau^{-}\rightarrow
\nu_{\tau}\pi^{-})\approx11.06\%$, $\cos^{2}\theta_{c}\approx0.95$,
$m_{\tau}=1.78GeV$, $m_{u}\approx\frac{1}{2}m_{d}\approx4MeV$,
$m_{s}=115MeV$, $m_{\eta}=548MeV$, $m_{\eta'}=957MeV$, and
$m_{\pi}\approx135MeV$ [22].

Using above given values of the relevant parameters, we present the
branching ratios $Br(\tau^{-}\rightarrow \mu^{-}\pi^{0})$,
$Br(\tau^{-}\rightarrow \mu^{-}\eta)$, and $Br(\tau^{-}\rightarrow
\mu^{-} \eta')$ as functions of the mixing factor $K$ for three
values of the mass $M_{\pi_{t}}$ in Fig.1, Fig.2, and Fig.3,
respectively. From these figures, we can see that the values of the
branching ratios $Br(\tau^{-}\rightarrow \mu^{-}\pi^{0})$,
$Br(\tau^{-}\rightarrow \mu^{-}\eta)$, and $Br(\tau^{-}\rightarrow
\mu^{-}\eta')$ increase as the mixing parameter $K$ increasing and
the mass parameter $M_{\pi_{t}}$ decreasing. However, in all of the
parameter space, the values of these branching ratios are much
smaller than the corresponding experimental upper limits given in
Table 1. Thus, we have to say that the possible signatures of the
neutral top-pion $\pi_{t}^{0}$ can not be detected via the $LFV$
process $\tau^{-}\rightarrow l^{-}P$ in the future experiments.

\noindent{\bf IV. The non-universal gauge boson $Z'$ and the $LFV$
semileptonic $\tau$ decays \hspace*{0.6cm} $\tau \rightarrow lP(V)$
}

The new physics models beyond the $SM$ generally predict the
existence of extra neutral gauge boson $Z'$  . If discovered it
would represent irrefutable proof of new physics, most likely that
the $SM$ gauge group should be extended [23]. If these extensions
are associated with flavor symmetry breaking, the gauge interactions
will not be flavor-universal which predict the existence of
non-universal gauge boson $Z'$ [16]. This kind of new particles can
lead to rich phenomenology [for review see [24]]. In this section,
we will consider the contributions of the non-universal gauge boson
$Z'$ predicted by the $TC2$ models to the $LFV$ semileptonic $\tau$
decays $\tau^{-}\rightarrow \mu^{-}P$
 and $\tau^{-}\rightarrow \mu^{-}V$.

Using the expressions of the effective four fermion couplings
$\tau\mu qq$ $(q=u,\ d,\ c,$ and $s)$ given in Eq.(4), the effective
interactions, which are related our calculation, can be written as:
\begin{eqnarray}
L^{eff}_{\tau\mu\pi}&=&F_{\pi}[A^{\pi}_{L}\bar{\tau}_{L}\gamma^{\mu}\mu_{L}
+A^{\pi}_{R}\bar{\tau}_{R}\gamma^{\mu}\mu_{R}]\partial_{\mu}\pi^{0}+h.\
c.\
,\\L^{eff}_{\tau\mu\eta}&=&F_{\eta}[A^{\eta}_{L}\bar{\tau}_{L}\gamma^{\mu}\mu_{L}
+A^{\eta}_{R}\bar{\tau}_{R}\gamma^{\mu}\mu_{R}]\partial_{\mu}\eta+h.\
c.\
,\\L^{eff}_{\tau\mu\rho}&=&{\frac{m^{2}_\rho}{g_{\rho}}}[A^{\rho}_{L}\bar{\tau}_{L}\gamma^{\mu}\mu_{L}
+A^{\rho}_{R}\bar{\tau}_{R}\gamma^{\mu}\mu_{R}]\rho^{0}_{\mu}+h.\
c.\
,\\L^{eff}_{\tau\mu\phi}&=&{\frac{m^{2}_\phi}{g'_{\phi}}}[A^{\phi}_{L}\bar{\tau}_{L}\gamma^{\mu}\mu_{L}
+A^{\phi}_{R}\bar{\tau}_{R}\gamma^{\mu}\mu_{R}]\phi^{0}+h.\ c.\
\\ with \nonumber\\ \
A^{\pi}_{L}&=&A_{L}^{\rho}=\frac{A}{2},\hspace{0.7cm}
A^{\pi}_{R}=A^{\rho}_{R}=A;
\\A^{\eta}_{L}&=&\frac{A}{2\sqrt{3}}, \
\hspace{1.1cm} A^{\eta}_{R}=\frac{A}{\sqrt{3}};
\\ A^{\phi}_{L}&=&\frac{A}{3}, \
\hspace{1.6cm} A^{\phi}_{R}=\frac{A}{3}.
\end{eqnarray}
Where $A=\frac{g_{1}^{2}K'\tan \theta'}{4M_{Z'}^{2}}$,
$\frac{1}{g_{\rho}}\approx0.2$, and $\frac{1}{g'_{\phi}}\approx0.25$
[21].

In the context of the $TC2$ models, the expressions of the
corresponding branching ratios induced by the non-universal gauge
boson $Z'$ can be written as:
\begin{eqnarray}
Br(\tau^{-}\rightarrow\mu^{-}\pi^{0})&=&\frac{5g_{1}^{6}K'^{2}}{1024G^{2}_{F}\pi
K_{1}M^{4}_{Z'}\cos^{2} \theta_{c}}Br(\tau^{-}\rightarrow
\nu_{\tau}\pi^{-}),\\
Br(\tau^{-}\rightarrow\mu^{-}\eta)&=&\frac{5g_{1}^{6}K'^{2}}{3072G^{2}_{F}\pi
K_{1}M^{4}_{Z'}\cos^{2}
\theta_{c}}(\frac{F_{\eta}}{F_{\pi}})^{2}(1-\frac{m^{2}_{\eta}}{m^{2}_{\tau}})^{2}Br(\tau^{-}\rightarrow
\nu_{\tau}\pi^{-}),
\\Br(\tau^{-}\rightarrow\mu^{-}\rho^{0})&=&\frac{5g_{1}^{6}K'^{2}}{1024G^{2}_{F}\pi
K_{1}M^{4}_{Z'}\cos^{2} \theta_{c}}Br(\tau^{-}\rightarrow
\nu_{\tau}\rho^{-}),\\
Br(\tau^{-}\rightarrow\mu^{-}\phi)&=&\frac{5g_{1}^{6}K'^{2}}{1152G^{2}_{F}\pi
K_{1}M^{4}_{Z'}\cos^{2}
\theta_{c}}(\frac{m_{\phi}}{F_{\pi}})^{2}(1-\frac{m^{2}_{\phi}}{m^{2}_{\tau}})^{2}
(1+\frac{2m_{\phi}^{2}}{m_{\tau}^{2}})^{2}\nonumber\\
 & &Br(\tau^{-}\rightarrow \nu_{\tau}\pi^{-}).
\end{eqnarray}
Where $Br(\tau^{-}\rightarrow \nu_{\tau}\rho^{-})\approx25\%$,
$m_{\phi}=1.019\ GeV$, and $G_{F}=1.166\times10^{-5}\ GeV^{-2}$
[22]. By replacing $F_{\eta}\rightarrow F_{\eta'}$ and
$m_{\eta}\rightarrow m_{\eta'}$, we can easily give the expression
of the branching ratio $Br(\tau^{-}\rightarrow\mu^{-}\eta')$ from
Eq.(19). However, because of the cancellation between the decay
constants for the singlet and octet components in the $\eta'$ meson,
there is $F_{\eta^{'}}\approx\frac{1}{3}F_{\eta}$, so the value of
$Br(\tau^{-}\rightarrow\mu^{-}\eta')$ is approximately smaller than
that of $Br(\tau^{-}\rightarrow\mu^{-}\eta)$ by one order of
magnitude. Thus, we have not given the expression for
$Br(\tau^{-}\rightarrow\mu^{-}\eta')$ in above equations.

It has been shown that vacuum tilting and the constraints from
$Z$-pole physics and $U(1)$ triviality require $K_{1}\leq1$ [10].
The lower limits on the $Z'$ mass $M_{Z'}$ can be obtained via
studying its effects on various observables, which have been
precisely measured in the present high energy collider experiments
[8]. For example, the lower bounds on $M_{Z'}$ can be obtained from
dijet and dilepton production in the Tevatron experiments [25] or
$B\bar{B}$ mixing [26]. However, these bounds are significantly
weaker than those from the precision electroweak data. Ref.[16] has
shown that, to fit the precision electroweak data, the $Z'$ mass
$M_{Z'}$ must be larger than $1TeV$. In our numerical estimation, we
will assume that the value of $M_{Z'}$ is in the range of $1\ TeV\
\sim2\ TeV$.

\begin{figure}[htb] \vspace{-0.5cm}
\begin{center}
\epsfig{file=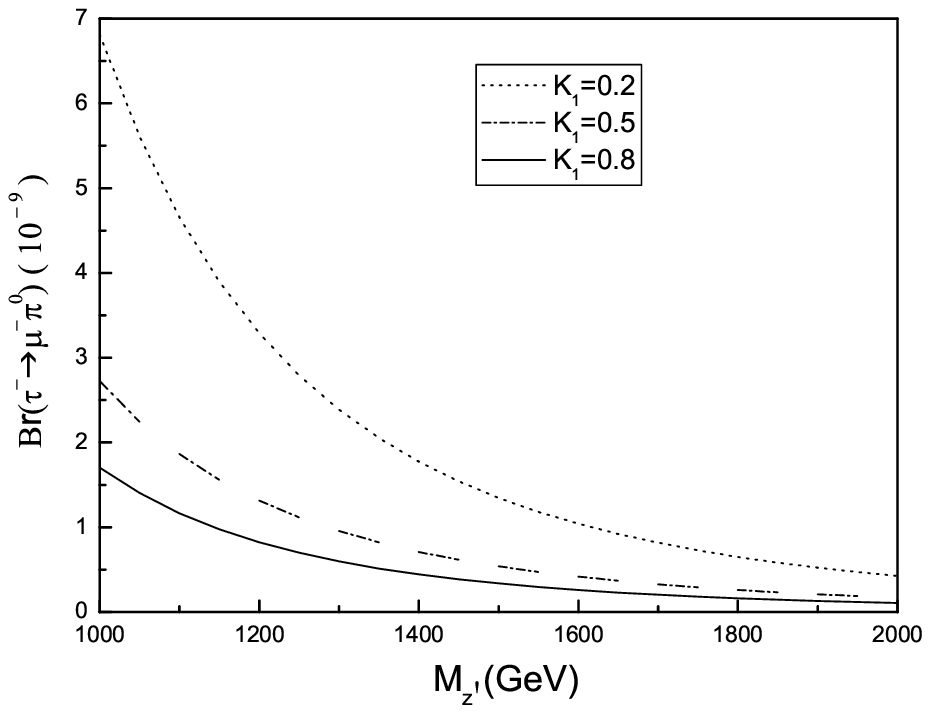,width=220pt,height=200pt} \put(-110,5){
(a)}\put(115,5){ (b)}
 \hspace{0cm}\vspace{-0.25cm}
\epsfig{file=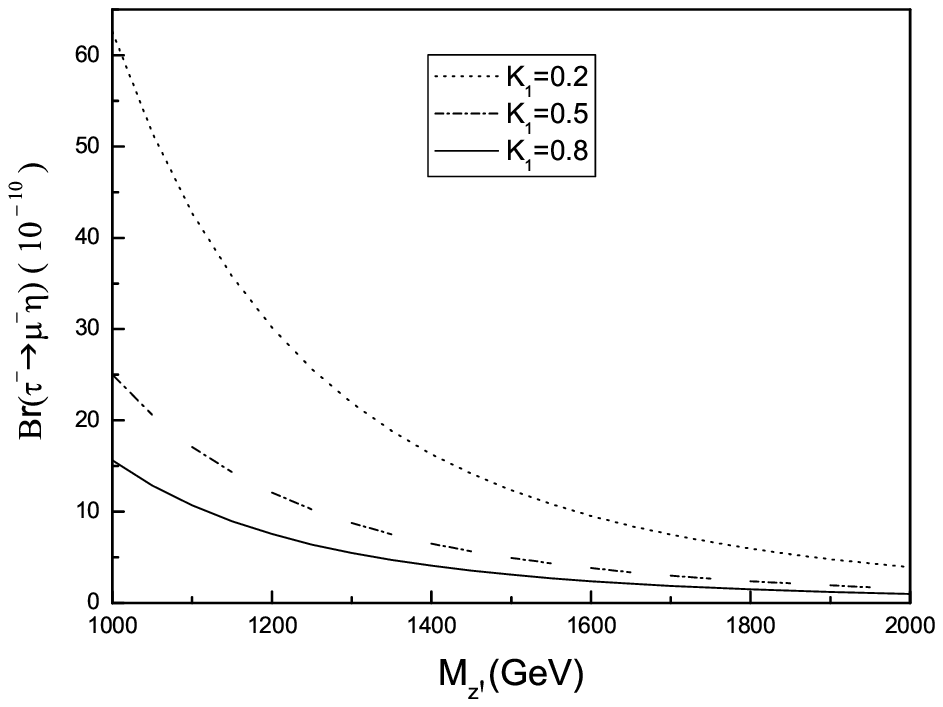,width=220pt,height=200pt} \hspace{-0.5cm}
 \hspace{10cm}\vspace{-1cm}
\epsfig{file=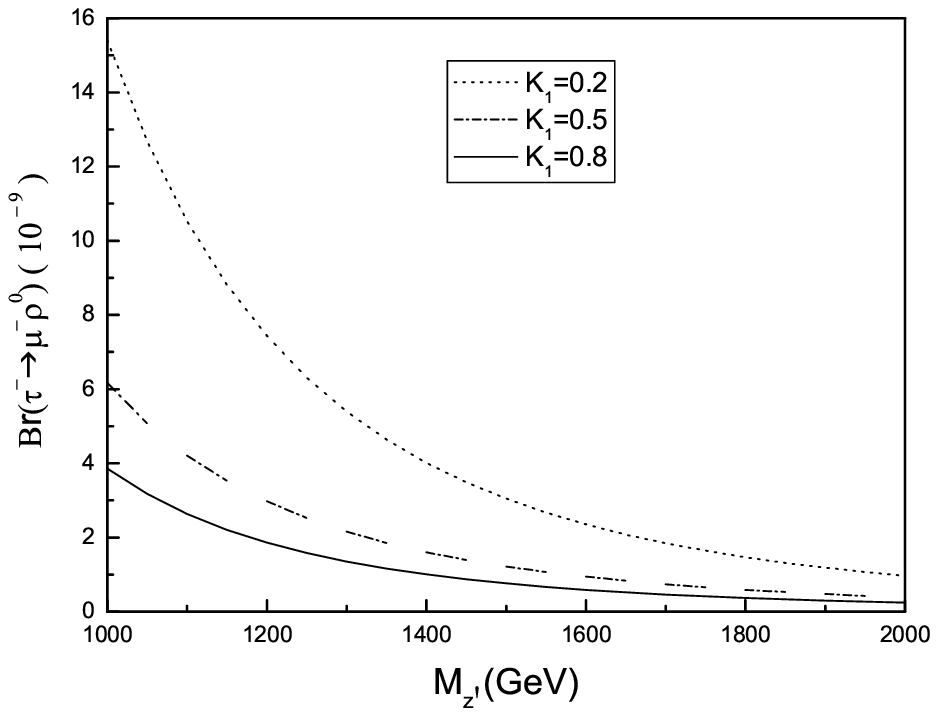,width=220pt,height=200pt} \hspace{-0.5cm}
 \hspace{0cm}\vspace{-0.25cm}
 \put(-110,5){(c)}\put(115,5){ (d)}
\epsfig{file=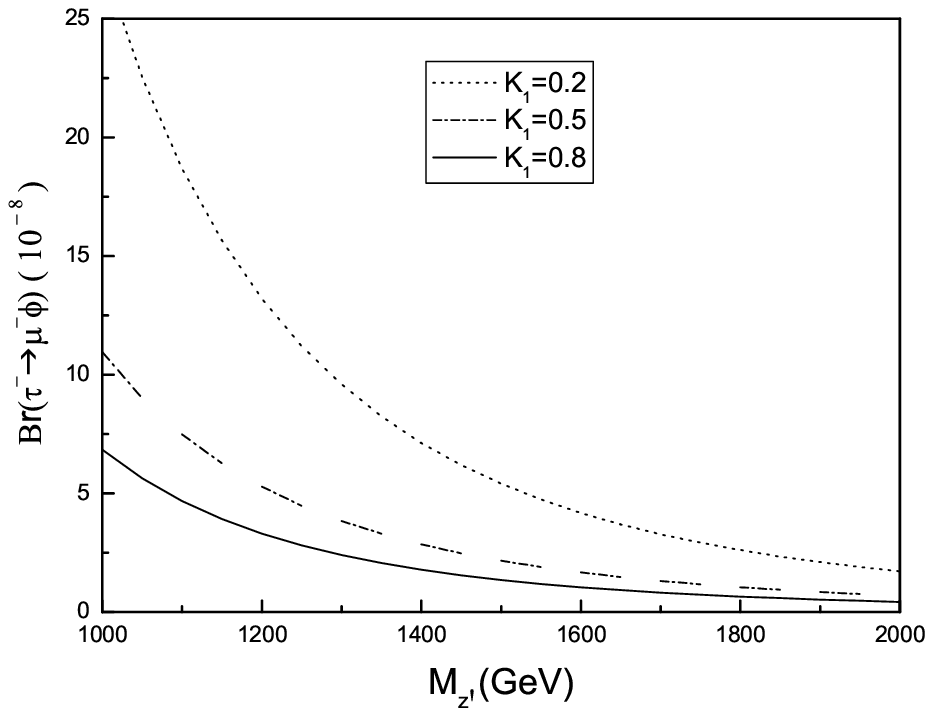,width=220pt,height=200pt} \hspace{-0.5cm}
 \hspace{10cm}\vspace{-0.2cm}
 \put(-200,-25) {Fig.4: The branching ratios as functions of the mass parameter
 $M_{Z'}$ for $K'=\frac{1}{\sqrt{2}}$ and } \put(-175,-40){three values of the
 coupling parameter $K_{1}$.}
\end{center}
\end{figure}
\vspace{-0.5cm}

The branching ratios $Br(\tau^{-}\rightarrow\mu^{-}P)$ and
$Br(\tau^{-}\rightarrow\mu^{-}V)$ contributed by the non-universal
gauge boson $Z'$ are  plotted as functions of the mass parameter
$M_{Z'}$ for $K'=\frac{1}{\sqrt{2}}$ and three values of the
coupling parameter $K_{1}$ in Fig.4a $\sim$ Fig.4d. From these
figures, one can see that the contributions of $Z'$ to the $LFV$
process $\tau^{-}\rightarrow \mu^{-}P$ are larger than those of the
neutral top-pion $\pi_{t}^{0}$ in most of the parameter space of the
$TC2$ models. For $K'=\frac{1}{\sqrt{2}}$, $M_{Z'}=1TeV$, and
$K_{1}=0.2$, the values of the branching ratios
$Br(\tau^{-}\rightarrow\mu^{-}\pi^{0})$ and
$Br(\tau^{-}\rightarrow\mu^{-}\eta)$ can reach 6.8$\times$10$^{-9}$
and 6.3$\times 10^{-9}$, respectively. However, these values are not
large enough to be detected in the future high energy experiments
[3]. For the $LFV$ processes $\tau^{-}\rightarrow\mu^{-}\rho^{0}$
and $\tau^{-}\rightarrow\mu^{-}\phi$, the values of the branching
ratios are larger than those for the $LFV$ processes
$\tau^{-}\rightarrow\mu^{-}\pi^{0}$ and
$\tau^{-}\rightarrow\mu^{-}\eta$. For $K'=\frac{1}{\sqrt{2}}$,
$0.1\leq K_{1}\leq 0.8$, and $1TeV\leq\ M_{Z'}\leq\  2TeV$, the
branching ratios $Br(\tau^{-}\rightarrow\mu^{-}\rho^{0})$ and
$Br(\tau^{-}\rightarrow\mu^{-}\phi)$ are in the ranges of 3.1$\times
10^{-8}\sim 2.4\times 10^{-10}$ and 5.5$\times10
^{-7}\sim4.3\times10^{-9}$, respectively. We expect that the value
of $Br(\tau^{-}\rightarrow\mu^{-}\phi)$ might approach the
corresponding experimental upper limits [3].

In above figures, we have taken the flavor mixing parameter $K'$ as
a fixed constant. In fact, for the $TC2$ models, the extended gauge
groups are broken at the $TeV$ scale, which proposes that $K'$ is an
$\cal{O}$$(1)$ free parameter. Its value can be generally
constrainted by the current experimental upper limits on the $LFV$
processes $l_{i}\rightarrow l_{j}\gamma$ and $l_{i}\rightarrow
l_{j}l_{k}l_{l}$. However, from the numerical results of Ref.[18],
we can see that the $LFV$ processes $l_{i}\rightarrow l_{j}\gamma$
and $l_{i}\rightarrow l_{j}l_{k}l_{l}$ can not give severe
constraints on the mixing factor $K'$. Thus, we expect that
$K'=\frac{1}{\sqrt{2}}$ is consistent with theoretically-allowed
parameter regions and also with current experimental data.

The non-universal gauge boson $Z'$ can also induce the effective
coupling $\tau^{-}\mu^{-}f\overline{f}$ via the off-shell photon
penguin diagrams, $i. \ e.$ the effective process
$\tau^{-}\rightarrow\mu^{-}\gamma^{\ast}\rightarrow\mu^{-}f\overline{f}$,
which can contribute to the $LFV$ semileptonic $\tau$ decay
processes $\tau^{-}\rightarrow\mu^{-}\rho^{0}$ and
$\tau^{-}\rightarrow\mu^{-}\phi$. However, the contributions of the
off-shell photon penguin diagrams induced by $Z'$ exchange to the
$\tau^{-}\mu^{-}f\overline{f}$ coupling are much smaller than those
of $Z'$ exchange at tree level[18]. Thus, in this paper, we have
neglected the contributions of the off-shell photon penguin diagrams
to the $LFV$ processes $\tau^{-}\rightarrow\mu^{-}\rho^{0}$ and
$\tau^{-}\rightarrow\mu^{-}\phi$.

\noindent{\bf V. Conclusions and discussions }

The evidence for the neutrino masses and flavor mixing, which can be
seen as the first experimental clue of new physics beyond the $SM$,
implies the non-conservation of the lepton flavor symmetry. Thus,
the $LFV$ processes in the charged lepton sector are expected, which
are very sensitive to new physics beyond the $SM$. Considering the
sensitivity of probing the $LFV$ semileptonic $\tau$ decays have
been enhanced to $\cal{O}$$(10^{-7})$, we calculate the branching
ratios for the $LFV$ processes $\tau^{-}\rightarrow\mu^{-}P$
$(P=\pi^{0},\ \eta,\ \eta'\ )$ and $\tau^{-}\rightarrow\mu^{-}V$
$(V=\rho^{0},\ \phi)$ in the context of the $TC2$ models.

A common feature of topcolor scenario is that it predicts the
existence of the neutral top-pion $\pi_{t}^{0}$ and the
non-universal gauge boson $Z'$, which have the tree-level $FC$
couplings to ordinary leptons. Thus, these new particles can
generate significant contributions to the $LFV$ processes. In this
paper, we have calculated the contributions of $\pi_{t}^{0}$ and
$Z'$ predicted by the $TC2$ models to the $LFV$ processes
$\tau^{-}\rightarrow l^{-}P$ and $\tau^{-}\rightarrow l^{-}V$. Our
numerical results show that, in most of the parameter space, the
neutral top-pion $\pi_{t}^{0}$ can only make the values of the
branching ratios $Br(\tau^{-}\rightarrow l^{-}\pi^{0})$ and
$Br(\tau^{-}\rightarrow l^{-}\eta(\eta'))$ in the range of
1$\times10^{-11}$ $\sim$ 1$\times10^{-16}$, which are still several
orders of magnitudes below the accessible current experimental
bounds. For the non-universal gauge boson $Z'$, its contributions to
the $LFV$ semileptonic $\tau$ decays are generally larger than those
of the neutral top-pion $\pi_{t}^{0}$. For example, with reasonable
values of the free parameters in the $TC2$ models, $Z'$ exchange can
make the value of the branching ratio $Br(\tau^{-}\rightarrow
\mu^{-}\phi)$ reach 1 $\times10^{-7}$, which might approach the
detectability threshold of near future experiments. Certainly, our
numerical results are strongly depend on the values of the mixing
parameter $K'$ and the mass parameter $M_{Z'}$.

Some popular models beyond the $SM$, such as $SUSY$, little Higgs
models, and extra dimension models, predict the existence of the
extra neutral gauge boson $Z'$, which generally has the $LFV$
coupling to leptons and might produce significant contributions to
the $LFV$ semileptonic $\tau$ decays. One can use these decay
processes to measure the coupling strength of $Z'$ with leptons and
to distinguish the topcolor models from other new physics models via
definition of an angular asymmetry [7]. More studying about the
effects of the extra gauge boson $Z'$ on the $LFV$ semileptonic
$\tau$ decays is needed and it will be helpful to discriminate
various specific models beyond the $SM$ in the future high energy
experiments.

\vspace{0.5cm} \noindent{\bf Acknowledgments}

C. X. Yue would like to thank the {\bf Abdus Salam } International
Centre for Theoretical Physics(ICTP) for partial support. This work
was supported in part by Program for New Century Excellent Talents
in University(NCET-04-0290), the National Natural Science Foundation
of China under the Grants No.10475037 and 10675057.

\null
\end{document}